# Direct growth of germanene at interfaces between van der Waals materials and Ag(111)

*Seiya Suzuki,* Takuya Iwasaki, K. Kanishka H. De Silva, Shigeru Suehara, Kenji Watanabe, Takashi Taniguchi, Satoshi Moriyama, Masamichi Yoshimura, Takashi Aizawa, and Tomonobu Nakayama*

Dr. S. Suzuki, Dr. T. Iwasaki, Dr. T. Taniguchi, Prof. T. Nakayama
International Center for Young Scientists (ICYS),
National Institute for Materials Science (NIMS)
1-1 Namiki, Tsukuba 305-0044, Japan
E-mail: SUZUKI.Seiya@nims.go.jp, seiya09417@gmail.com

Dr. T. Iwasaki, Dr. T. Taniguchi, Dr. S. Suehara, Dr. S. Moriyama,[+] Dr. T. Aizawa, Prof. T. Nakayama
International Center for Materials Nanoarchitectonics (WPI-MANA),
National Institute for Materials Science (NIMS)
1-1 Namiki, Tsukuba 305-0044, Japan

Dr. K. K. H. De Silva, Prof. M. Yoshimura
Toyota Technological Institute
2-12-1 Hisakata, Tempaku, Nagoya 468-8511, Japan,

Dr. K. Watanabe
Research Center for Functional Materials,
National Institute for Materials Science (NIMS)
1-1 Namiki, Tsukuba 305-0044, Japan

Prof. T. Nakayama
Graduate School of Pure and Applied Sciences,
University of Tsukuba
1-1 Namiki, Tsukuba 305-0044, Japan

[+]Present address: School of Engineering, Tokyo Denki University, 5 Senju-Asahi-cho, Adachi-ku, Tokyo 120-8551, Japan

Abstract

Germanene, a two-dimensional honeycomb germanium crystal, is grown at graphene/Ag(111) and hexagonal boron nitride (h-BN)/Ag(111) interfaces by segregating germanium atoms. A simple annealing process in $N_2$ or $H_2$/Ar at ambient pressure leads to the formation of germanene, indicating that an ultrahigh-vacuum condition is not necessary. The grown germanene is stable in air and uniform over the entire area covered with a van der Waals (vdW) material. As an important finding, it is necessary to use a vdW material as a cap layer for the present germanene growth method since the use of an $Al_2O_3$ cap layer resulted in no germanene formation. The present study also proved that Raman spectroscopy in air is a powerful tool for characterizing germanene at the interfaces, which is concluded by multiple analyses including first-principles density functional theory calculations. The direct growth of h-BN-capped germanene on Ag(111), which is demonstrated in the present study, is considered to be a promising technique for the fabrication of future germanene-based electronic devices.

## 1. Introduction

Two-dimensional honeycomb lattices of group IV (group 14) elements, such as silicene,[1] germanene,[1c, 2] stanene,[3] and plumbene,[4] are known as Xenes[5]. On the basis of theoretical predictions, Xenes have electronic properties similar to those of graphene such as a Dirac cone with linear band dispersion, leading to an extremely high carrier mobility.[6] In contrast to graphene, the crystal structure of Xenes has been suggested to be not flat but buckled. The buckled Xenes are expected to have a bandgap that can be tuned by applying a vertical electric field,[6] which can overcome the problem of gapless graphene for future electronic device applications. Thus, the utilization of Xenes in electronics is highly desirable.

Xenes have been grown on various crystal surfaces. Germanene has been grown on Ag(111),[7] Au(111),[2, 8] Cu(111),[9] Al(111),[10] Pt(111),[11] graphite,[12] and $MoS_2$,[13] and silicene has been grown on Ag(111),[1a, 1b] Ag(110),[14] Al(111),[15] Ir(111),[16] $ZrB_2$(0001),[17] and ZrC(111).[18] A limited number of reports of stanene growth on Ag(111)[3b] and Cu(111),[3a] as well as one report of plumbene growth on a $Pd_{1-x}Pb_x$(111) alloy film on Pd(111),[4a] are available. Although the growth of Xenes has been widely reported, Xenes-based electronic devices have not been reported, except for one silicene field-effect transistor (FET).[19] One of the reasons for this is the chemical instability of Xenes, in contrast to the stability of graphene.[20] Although high-quality Xenes can be grown in ultrahigh vacuum (UHV), they are expected to be immediately oxidized when removed from the UHV chamber.[21] Hence, for the realization of Xenes-based applications, we have to overcome this problem.

As one approach, we conceived the direct growth of Xenes at interfaces. The oxidation of a Xene should be prevented by placing it at an interface that provides a spatial separation between Xenes and oxidizing substances in air. A material with excellent gas barrier properties is required to form such an interface. In this paper, we report a novel method of growing germanene at interfaces between van der Waals (vdW) materials and Ag(111). Since vdW materials, such as graphene and hexagonal boron nitride (h-BN), have excellent gas barrier properties,[22] they were used as a cap layer to form an interface for germanene growth. The proposed method of growing germanene is schematically drawn in **Figure 1**. First, graphene or h-BN was transferred onto a Ag(111) thin film formed on Ge(111). To induce Ge segregation at the vdW material/Ag(111) interface, the sample was annealed in $N_2$ or $H_2$/Ar ($H_2$ concentration ~ 3%) ambient at atmospheric pressure. To characterize the germanene at the interface, Raman spectroscopy in air and first-principles density functional theory (DFT) and density functional perturbation theory (DFPT) calculations were performed, and the two Raman peaks observed at ~155 and ~255 $cm^{-1}$ were assigned as out-of-plane and in-plane vibration modes of germanene, respectively. As a result, it was revealed that germanene was grown uniformly under the vdW material and was stable in air as expected. One of our key findings is that the germanene growth method requires a vdW material as a cap layer since the use of an $Al_2O_3$ cap layer resulted in the absence of germanene formation. To the best of our knowledge, this is the first report of the growth of a Xene at ambient pressure, which is beneficial from the viewpoint of reducing the cost of mass-producing future electronic devices based on germanene, as an alternative to conventional UHV growth.

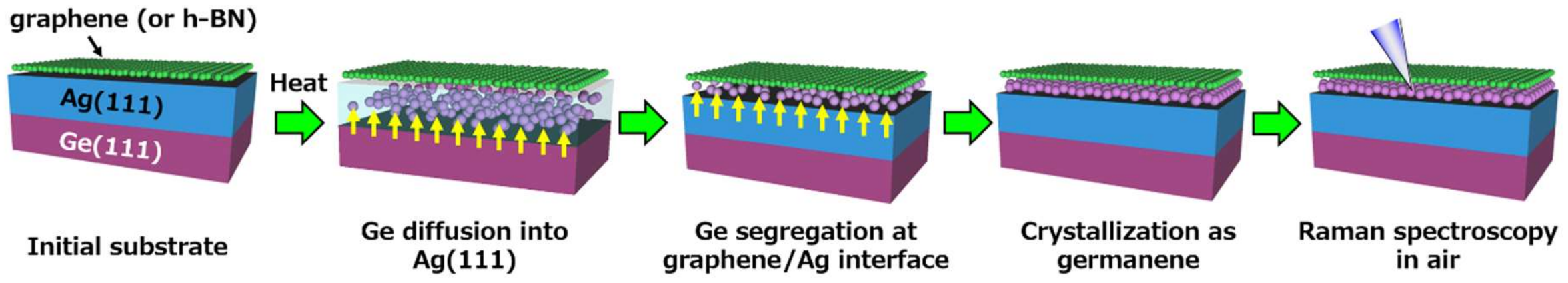

**Figure 1.** Illustration of germanene growth at vdW material/Ag (111) interface and its characterization by Raman spectroscopy in air.

## 2. Results and Discussion

**Figure 2** shows Raman spectra of graphene/Ag(111)/Ge(111) after annealing at room temperature (RT), 550, and 700 ºC, and the Raman spectrum of Ag(111)/Ge(111) without top graphene after annealing at 550 ºC. The top graphene was placed on the Ag surface by the wet transfer of chemical vapor deposition (CVD) graphene. It was found that two new peaks appeared around 155 and 255 $cm^{-1}$ after heating at 550 ºC (Figure 2b) as well as after heating at 450, 500, and 600 ºC (**Figure S1**). The two peaks did not appear in the case without the top graphene (Figure 2c). On the other hand, a sharp peak at 300 $cm^{-1}$, which corresponds to the Ge-Ge vibration mode of bulk Ge, was observed in the sample heated to 700 ºC (Figure 2d). This was caused by the surface exposure of Ge(111) due to the disappearance of the Ag layer.[23] The disappearance of the Ag layer was due to its phase transition to a liquid and, as a result, most of the Ag diffused into the bulk Ge. This is reasonable because 700 ºC is sufficiently higher than the eutectic melting point of the Ag-Ge system (~ 650 ºC).[24] Note that the Raman peaks observed around 155 and 255 $cm^{-1}$ are different from the Ge-Ge vibration mode observed in bulk Ge.

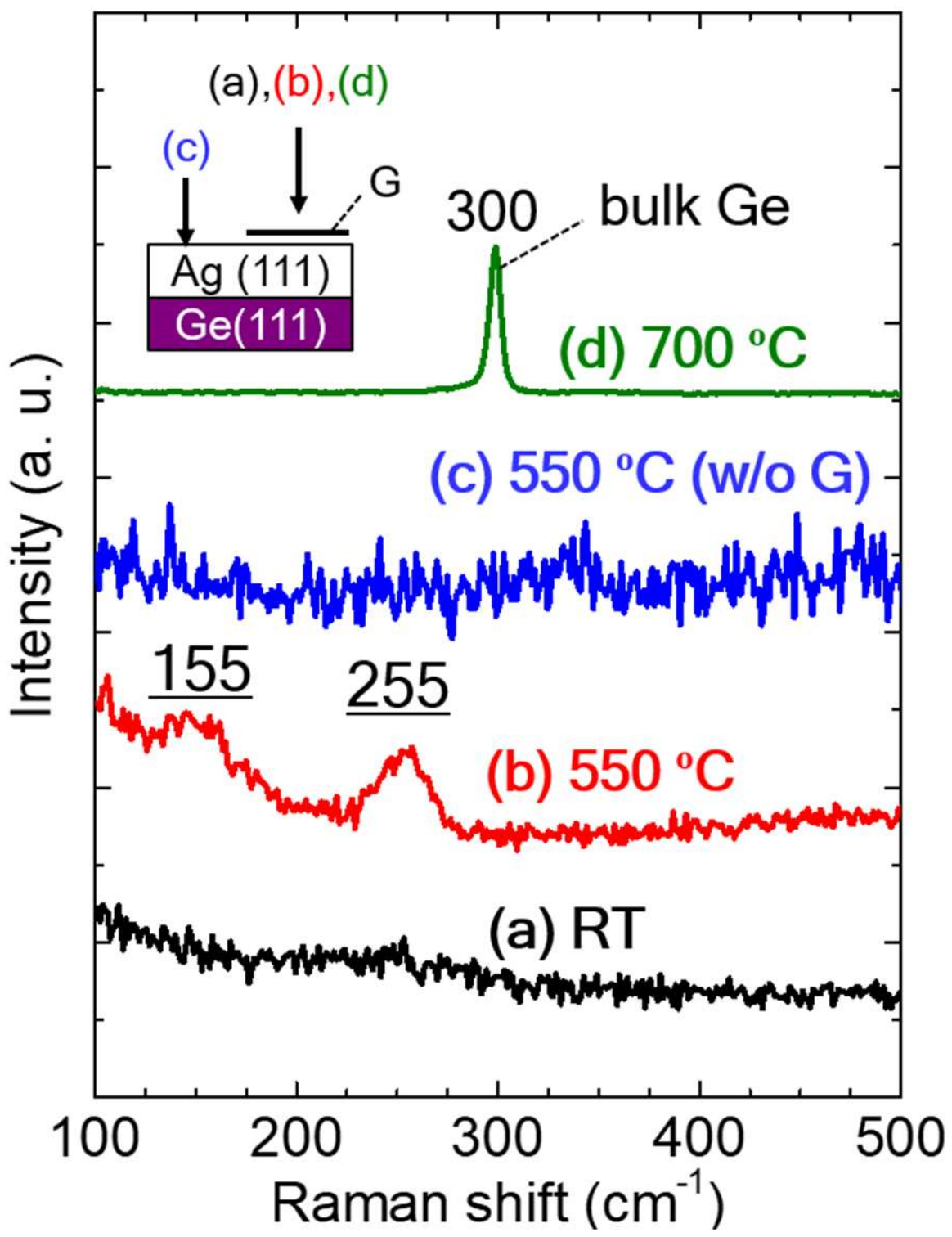

**Figure 2**. Raman spectra of graphene(G)/Ag(111)/Ge(111) after annealing at (a) RT, (b) 550, and (d) 700 ºC, and (c) Raman spectrum of Ag(111)/Ge(111) without graphene after annealing at 550 ºC. The top graphene was placed onto the Ag surface by the wet transfer of CVD graphene. The inset shows the areas where the spectra were obtained.

The graphene/Ag(111)/Ge(111) sample annealed at 550 ºC with two new Raman peaks was characterized by angle-resolved X-ray photoelectron spectroscopy (XPS) to reveal the distribution of elements in the vicinity of its surface. **Figure 3**a shows the XPS spectrum obtained at a photoelectron detection angle of 45º. Carbon, sulfur, oxygen, germanium, and silver were detected. Figure 3b shows the atomic concentration of each element as a function of the photoelectron detection angle. The concentrations of carbon, sulfur, and oxygen increased and that of silver decreased with increasing detection angle. The maximum concentration of Ge was observed at 45º. Since XPS with a larger detection angle is more sensitive to the surface, the mountain-like behavior of the Ge atomic concentration graph (Figure 3b) indicates that Ge atoms were located at the interface between graphene and Ag(111). The detected sulfur was thought to be mainly derived from impurities adsorbed on the Ag surface before the transfer of graphene. Figure 3c shows the Ge 3d peak obtained at 45° with high energy resolution. Elemental Ge is the main component (29. 6 eV) and the minor peak originates from oxides of Ge (32.6 eV). This oxide formation may have been due to oxidation in the region uncovered with graphene (where bare Ge surface was exposed) and the diffusion of oxygen from graphene edges or defects. These results suggest that a crystal composed of elemental Ge that exhibits the newly observed Raman peaks can exist at the interface between graphene and Ag(111).

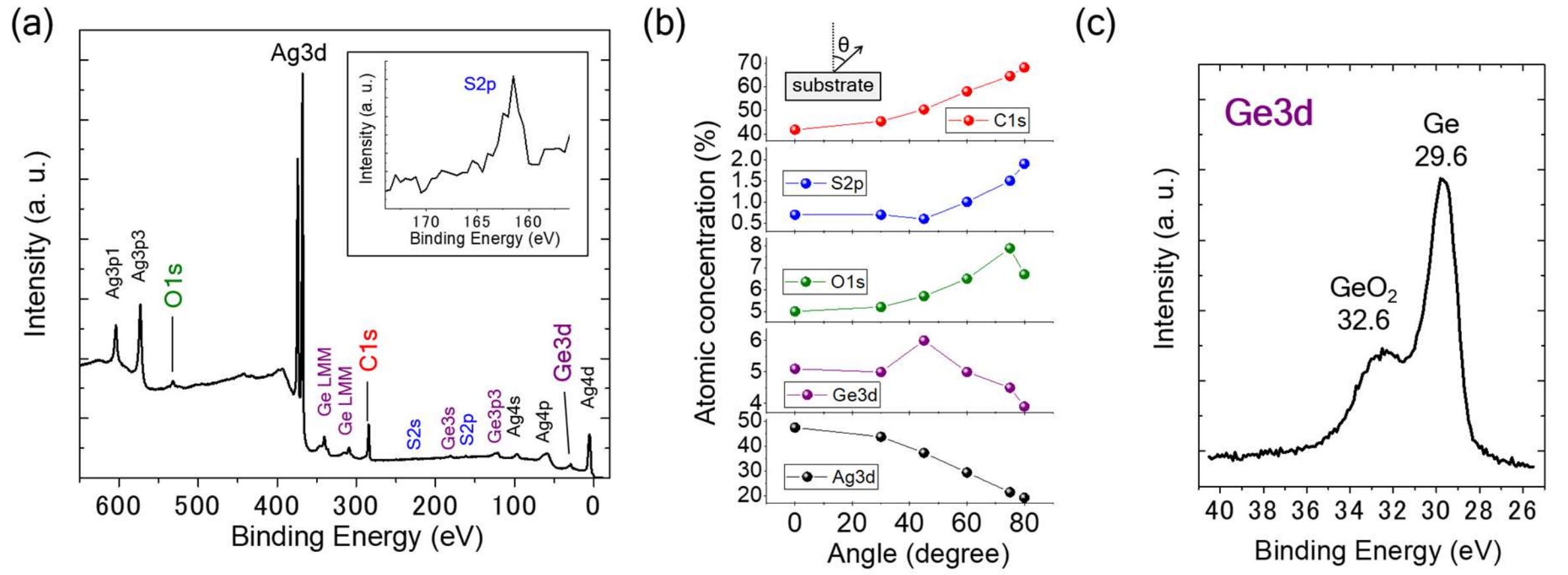

**Figure 3.** (a) XPS spectrum of graphene/Ag(111)/Ge(111) annealed at 550 ºC. (b) Atomic concentration of each element as a function of the photoelectron detection angle. The inset shows the definition of the detection angle. (c) Ge 3d peak obtained at 45° with high energy resolution. The annealed sample was kept in air for more than 42 days before introducing it into the XPS chamber.

To assign the new Raman peaks at 155 and 255 $cm^{-1}$ (Figure 2b), first-principles DFPT calculations were performed. We only focused on monolayer germanene since cross-sectional scanning transmission electron microscopy (STEM) observation indicated that the segregated

Ge at the interface is likely to be monolayer not multilayer (**Figure S2**). **Figure 4**a and Figure 4b show the calculated phonon dispersions of bulk Ge and freestanding germanene, respectively, and Figure 4c shows an illustration of the Raman active vibration modes in germanene obtained by vibration analysis. Raman active phonons are basically located at the gamma point, except in special Raman processes such as double resonance Raman scattering in graphene.[25] Thus, bulk Ge and germanene have one and two Raman active phonons, respectively, which has also been reported elsewhere.[26] This feature explains the experimental Raman spectra in Figure 2. Bulk Ge has only one peak at 300 $cm^{-1}$ (Figure 2d), while germanene has two peaks at 155 and 255 $cm^{-1}$ for out-of-plane and in-plane vibration modes, respectively.

The experimentally observed phonon for the in-plane mode of germanene was ~ 45 $cm^{-1}$ lower than that of bulk Ge. It seems that the experimentally observed germanene peaks are redshifted from the freestanding germanene peaks since the calculated phonon energies of bulk Ge and the in-plane mode in freestanding germanene are almost the same (Figure 4a and Figure 4b). The energy shift of the phonon is caused by the change in bond length, which is generally induced by strain and charge transfer to the lattice. In the present case, rather than the top graphene, Ag(111) is considered to contribute to phonon energy shifts since the naked Ag(111) surface is chemically reactive.

To consider the redshift, we first examined the effect of strain on freestanding germanene. Figure 4d shows the phonon energies of the in-plane and out-of-plane modes and the Ge-Ge bond length in freestanding germanene as a function of in-plane biaxial strain. It is clear that tensile strain reduces the phonon energies for both modes and increases the Ge-Ge bond length. Subsequently, we calculated a possible lattice structure of germanene on a Ag(111) (7 × 7) surface. The DFT calculations yield buckled hexagonal germanene on a Ag(111) (7 × 7) surface as shown in the top and side views in Figure 4e and Figure 4f, respectively. By extracting the coordinates of Ge atoms, the average Ge-Ge bond length was obtained as ~ 2.49 Å. which corresponds to slightly tensile freestanding germanene (~ 3%). Since the corresponding phonon energies are ~165 and ~ 260 $cm^{-1}$ for the out-of-plane and in-plane modes, respectively, the strain induced by the Ag(111) surface is mainly responsible for the redshift. Although we cannot yet give a complete quantitative discussion, the above explanation may provide helpful ideas for explaining the observed Raman peaks of germanene at the graphene/Ag(111) interface.

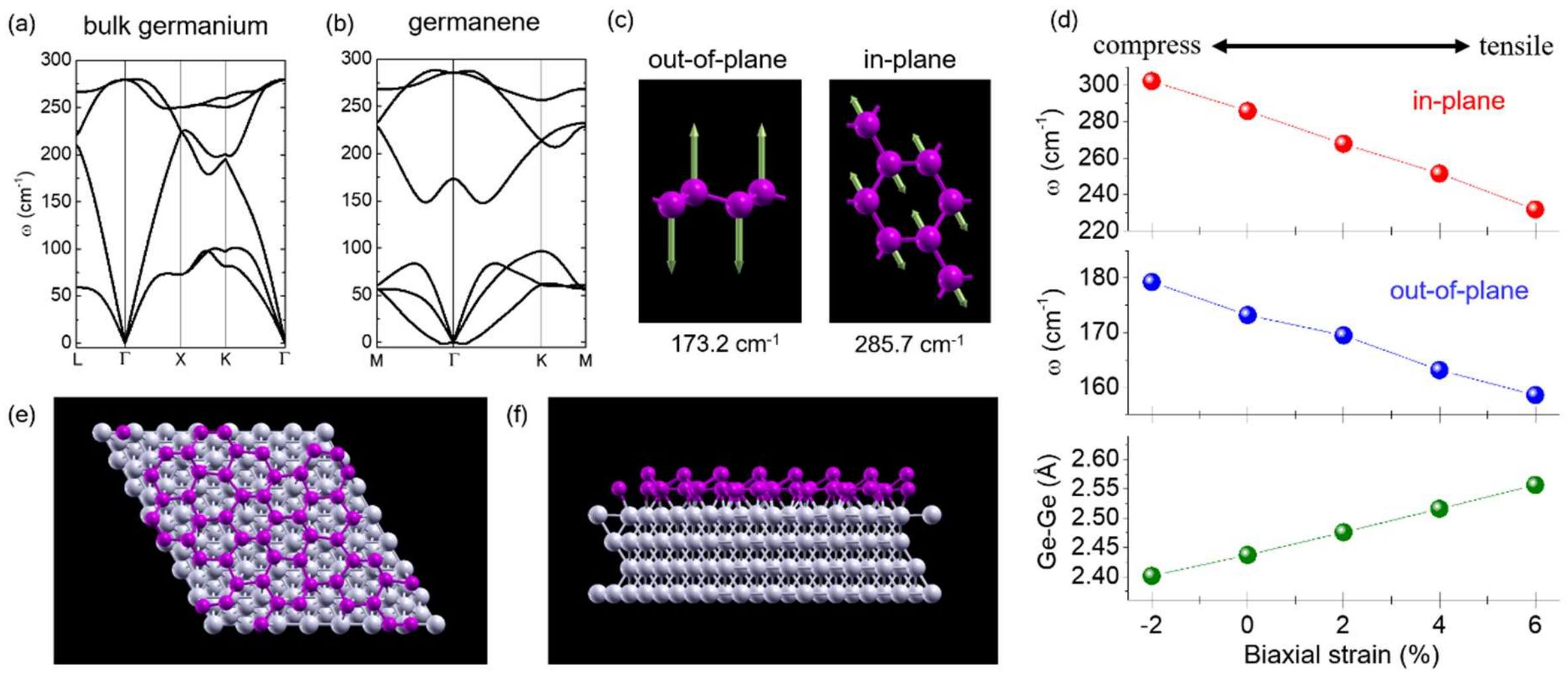

**Figure 4.** DFT calculations. Phonon dispersions of (a) bulk Ge and (b) freestanding germanene. (c) Illustration of Raman active vibration modes in germanene. (d) Phonon energies of in-plane and out-of-plane modes and the Ge-Ge bond length in freestanding germanene as a function of in-plane biaxial strain. (e) Top and (f) side views of the calculated germanene structure on four-layer Ag(111). Structural optimization was carried out with movable Ge and Ag atoms in the first surface layer and fixed Ag bulk positional atoms in the second to fourth surface layers. The average Ge-Ge bond length in germanene on Ag(111) was ~ 2.49 Å.

To double-check that the observed Raman peaks were really derived from germanene, we grew germanene in a UHV chamber using the well-known surface segregation method reported by Yuhara *et al*.[7a] The phonon energies of germanene were measured by *in situ* high-resolution electron energy-loss spectroscopy (HREELS) and *ex situ* Raman spectroscopy as illustrated in **Figure 5**a. First, Ge was cleaned by $Kr^+$ sputtering and annealing at 700 ºC, resulting in a c(2

× 8) reconstructed surface.[27] Ag was deposited at RT on the cleaned Ge by vacuum evaporation to form a Ag(111) epitaxial layer. Germanene surface segregation was induced by heating at 450 ºC for 10 min. The HREELS of germanene was performed in UHV to exclude the effects of impurities, such as oxygen, sulfur, and carbon. Subsequently, an amorphous boron (a-B) thin film was deposited at RT to protect the germanene from oxidation in air. Noted that the a-B deposition is our original idea, which has not been included in Ref. 12. The thickness of the a-B film was estimated to be about 2 nm from the signal decay of Ag and Ge observed by Auger electron spectroscopy (AES) upon a-B deposition. After that, the sample was removed from the UHV chamber and examined by Raman spectroscopy in air. The Raman spectra were recorded within 3 h of removing the sample from the UHV chamber.

Figure 5b shows the AES spectrum after the segregation growth of germanene. Only Ag and Ge were detected on the surface, indicating an impurity-free surface. Figure 5c and Figure 5d show reflection high-energy electron diffraction (RHEED) patterns of Ag(111) and germanene/Ag(111), respectively, with the $[11\bar{2}0]$ azimuth. Two types of streaks (inner and outer) appeared in the RHEED pattern of germanene (Figure 5d), in addition to the fundamental streaks of Ag(111) (Figure 5c). The inner streak (indicated by the white arrows in Figure 5d) corresponds to germanene with a lattice constant of ~ 3.89 Å at the rotation angle of 30º, which is in good agreement with that reported for germanene on Ag(111).[7a] The outer faint streak is caused by multiple diffractions by the Ag(111) lattice and germanene.

Figure 5e shows the Raman spectrum of the boron-coated germanene on Ag(111) recorded in air. Two peaks similar to those for germanene at the graphene/Ag(111) interface (Figure 2b) were observed, and these peaks were slightly blueshifted. These results provide further evidence that the Raman peaks in Figure 2b are derived from germanene. Note that we did not obtain any Raman peaks of germanene without the a-B layer in air.

To consider the aforementioned blueshift, we compared the Raman results with the off-specular HREEL spectrum of germanene on Ag(111) as shown in Figure 5f. Black (B) and red

(G) dotted lines in Figure 5f show the phonon energies of germanene at the graphene/Ag(111) and boron/Ag(111) interfaces, respectively, which were obtained by Raman spectroscopy. The positions of the red dotted lines (G) match the peak at ~32.5 meV (262 $cm^{-1}$) and the shoulder peak at ~19.5 meV (157 $cm^{-1}$) in the HREELS spectrum, indicating that the phonon energies of surface germanene are almost the same as those of germanene at the graphene/Ag(111) interface rather than boron-coated germanene. This implies that the graphene layer does not disturb the structure of germanene on Ag(111). The a-B layer slightly interacted with germanene, resulting in the blueshift in Figure 5e. The limited interaction of graphene with germanene enabled the segregation growth by the simple heating of graphene/Ag(111)/Ge(111) samples at atmospheric pressure.

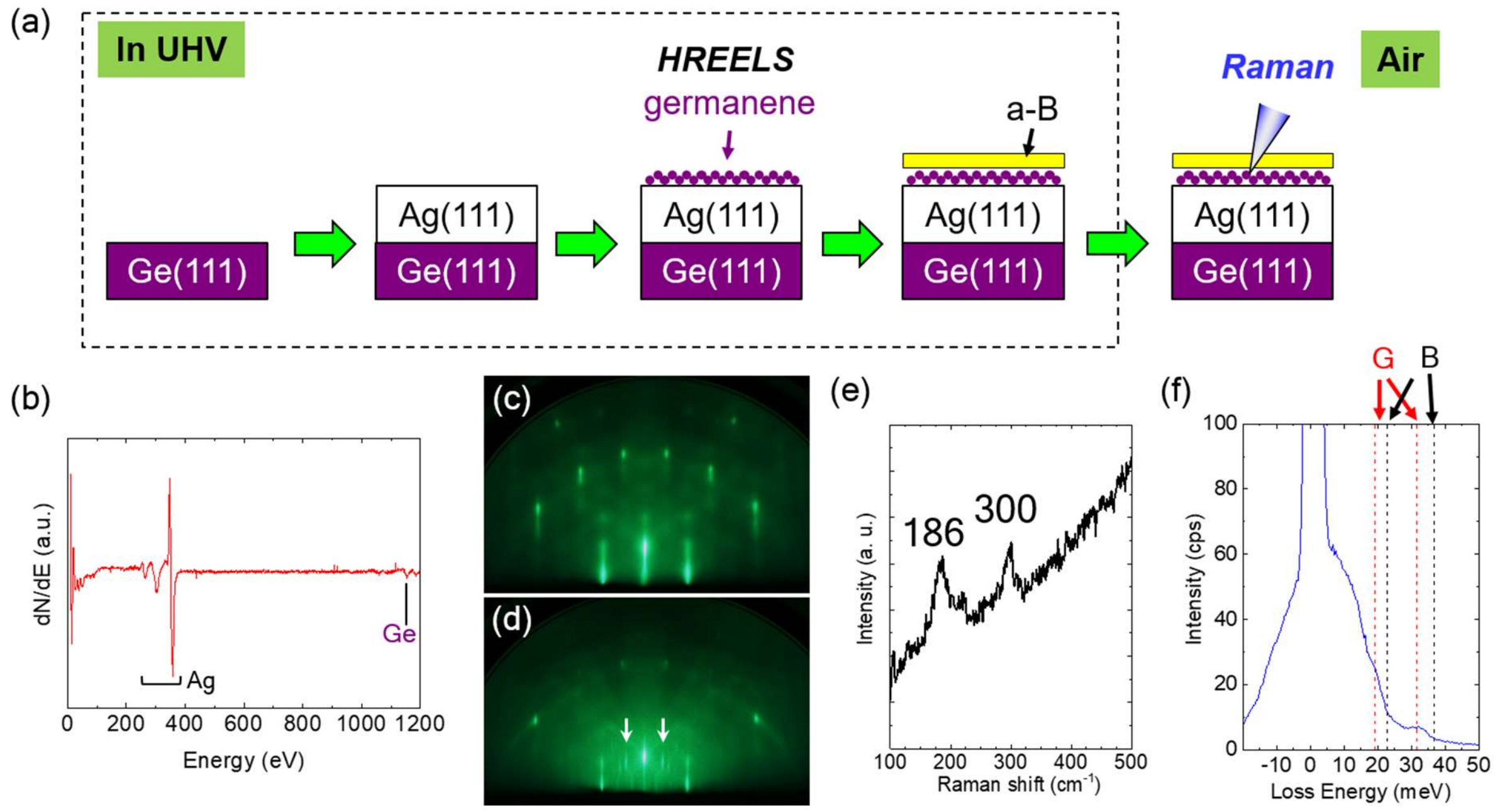

**Figure 5.** *In situ* growth of germanene. (a) Illustration of experimental procedure. (b) AES spectrum of germanene on Ag(111). RHEED patterns of (c) Ag(111) and (d) germanene/Ag(111) surface. Streaks with arrows correspond to germanene. The azimuth of the incident e-beam is $[11\bar{2}0]$. (e) Raman spectrum of boron-coated germanene on Ag(111) recorded in air. (f) Off specular HREEL spectrum of germanene on Ag(111). Black (B) and red

(G) dotted lines show the phonon energies of germanene at the graphene/Ag(111) and boron/Ag(111) interfaces, respectively, which were obtained by Raman spectroscopy.

We analyzed the uniformity of germanene at the graphene/Ag(111) interface by Raman mapping. **Figure 6**a shows an optical microscopy image of the sample, and Figure 6b and Figure 6c show its Raman maps of the full width at half maximum (FWHM) of the G' peak (graphene) and the peak intensity at 255 $cm^{-1}$ (germanene), respectively. The germanene peaks were observed throughout the graphene-covered region, as schematically shown in Figure 6d, indicating that a large area of germanene with high uniformity can be grown by the present method.

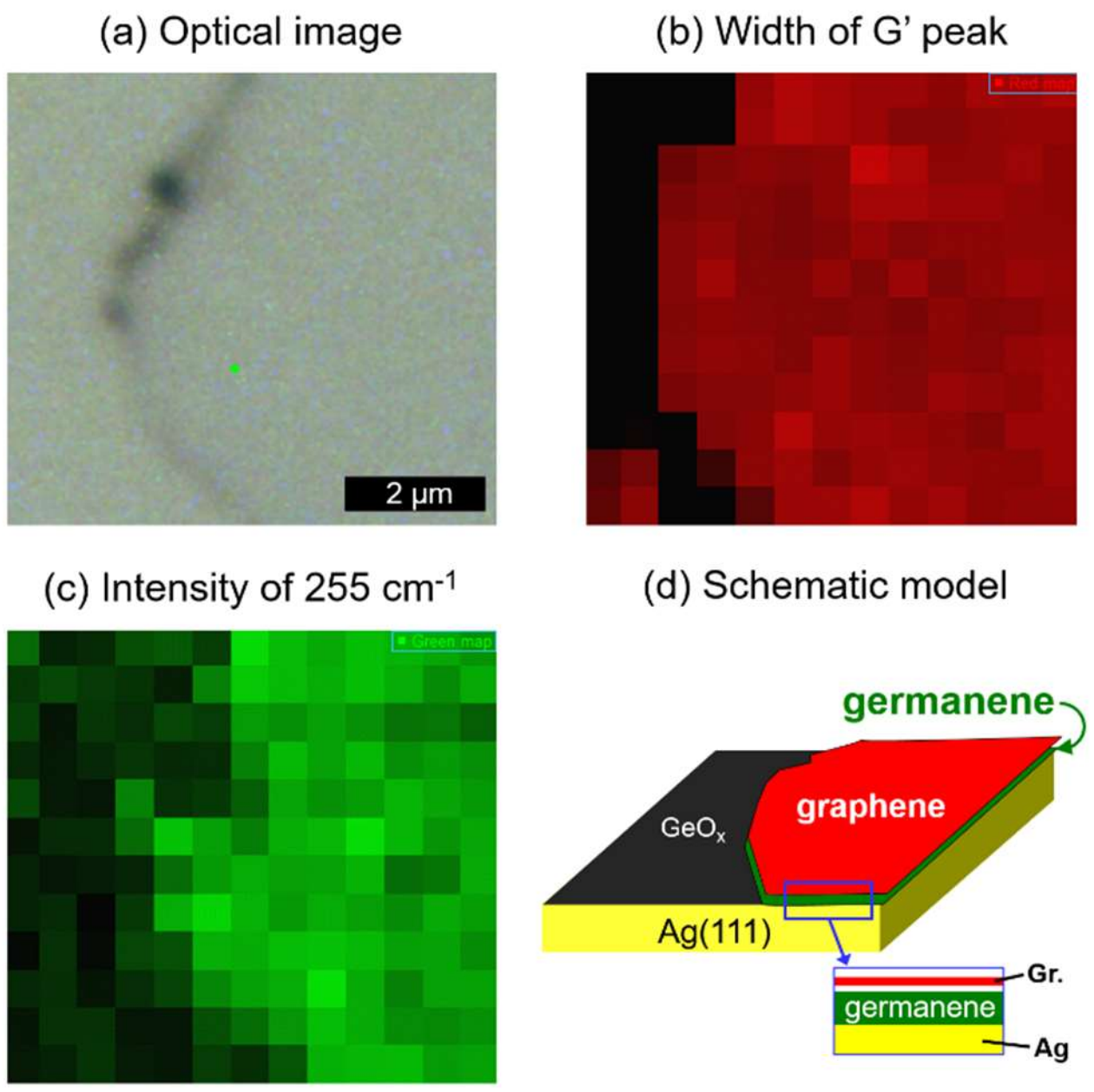

**Figure 6.** Raman mapping of germanene at graphene/Ag(111) interface. (a) Optical microscopy image of the sample. Raman maps for (b) full width at half maximum (FWHM) of the G' peak (graphene) and (c) peak intensity at 255 $cm^{-1}$ (germanene). (d) Schematic model of the Raman-mapped area of the sample.

To verify the possibility of using other types of cap layers, we tested multilayer exfoliated h-BN (~ 30 nm thickness), exfoliated graphene (monolayer) stacked with h-BN, and sputtered $Al_2O_3$ (~20 nm thickness) for the segregation growth of germanene from Ag(111)/Ge(111). **Figure 7** shows Raman spectra of the samples with the different cap layers after annealing at 550 ºC in $N_2$. Raman peaks of germanene were observed in the samples with all the graphene and h-BN cap layers, which are vdW materials, while no peaks were observed for the $Al_2O_3$ cap layer. This indicates that germanene were not formed at the interface between $Al_2O_3$ and Ag(111). For the formation of germanene, it is required for Ge atoms to segregate at the interface and two-dimensionally crystallize as germanene. Kurosawa et al. reported that Ge segregated at the $Al_2O_3$/Ag(111) interface and the segregated Ge formed Ge-O bonds when $Al_2O_3$/Ag(111)/Ge(111) was heated in $N_2$.[28] This indicates that the oxide interface is too reactive with germanene and thus prevents germanene formation at the interface. In contrast, the vdW interfaces, such as graphene and h-BN, are significantly less reactive with germanene, providing suitable conditions for growing germanene using the present method.

In addition, an h-BN cap layer is beneficial for the fabrication of germanene-based devices. As Tao *et al.* reported for the fabrication of silicene FETs,[19] germanene can also be transferred onto an insulating substrate. In this case, the h-BN cap layer becomes the top protective layer of germanene FETs. Since h-BN is the most suitable substrate for graphene FETs in terms of its low carrier scattering by impurities,[29] h-BN-capped germanene has great potential in terms of electronic properties, such as the tuning of its bandgap by an E-field and a high carrier mobility comparable to that of graphene.[30]

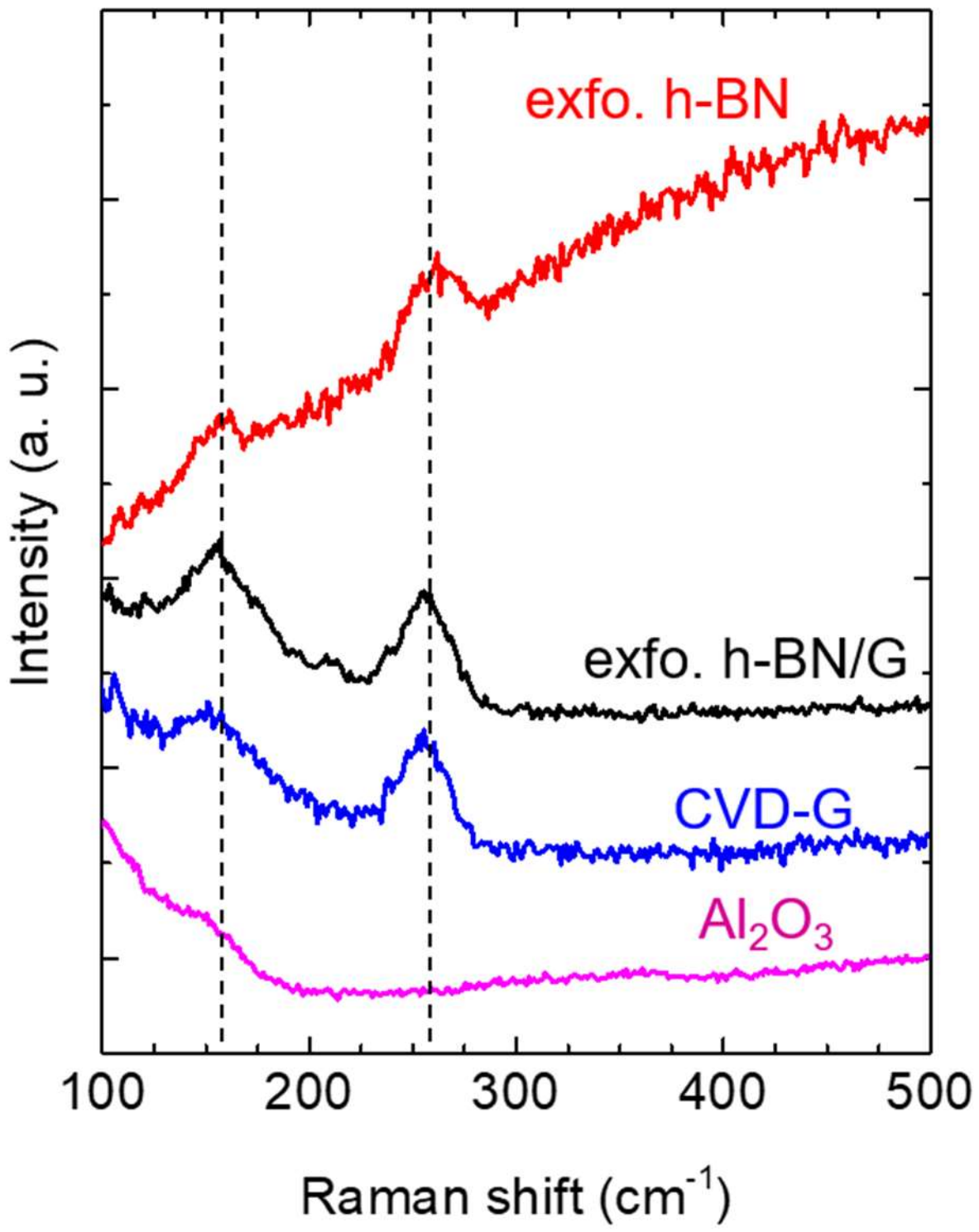

**Figure 7.** Raman spectra of germanene with the different cap layers. Multilayer exfoliated h-BN (~ 30 nm thickness), exfoliated graphene (monolayer) stacked with h-BN, CVD graphene (monolayer), and sputtered $Al_2O_3$ (~20 nm thickness) were deposited on Ag(111)/Ge(111). The annealing conditions were 550 ºC under $N_2$ ambient.

The stability of germanene at the h-BN/Ag(111) interface was also examined. **Figure 8**a shows Raman spectra of germanene at the interface after keeping the sample in air for 4, 39, and 55 days. The peak shape was preserved after 55 days in air. Figure 8b and Figure 8c show the extracted peak positions and FWHM of the out-of-plane and in-plane vibration modes, respectively. There are no significant changes in the features of the peaks, indicating that germanene at a vdW interface is stable in air.

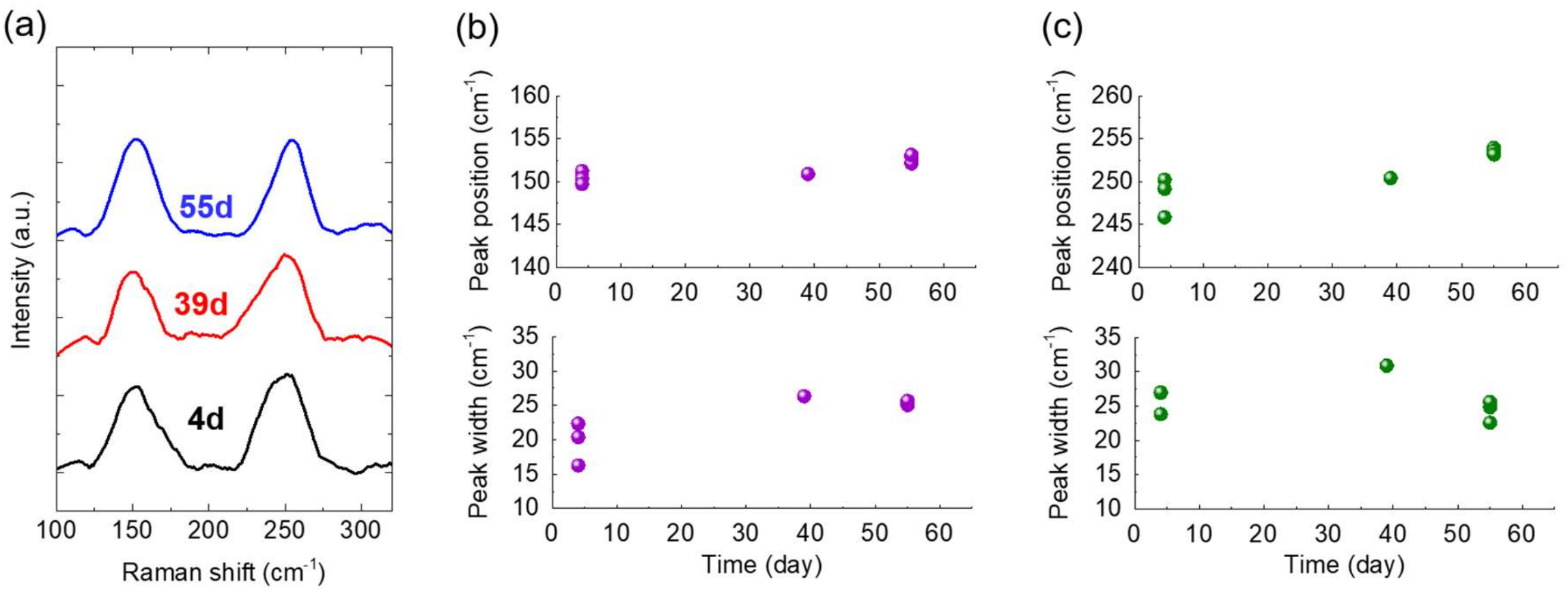

**Figure 8.** (a) Raman spectra of germanene at the h-BN/Ag(111) interface after keeping it in air for 4, 39, and 55 days. Smoothing and subtraction of baselines have been performed for the spectra, and then the peaks were fitted by a Lorentzian curve. Extracted peak positions and FWHM of (b) out-of-plane and (c) in-plane vibration modes of germanene.

## 3. Conclusion

We have uniformly grown large-area germanene at a vdW material/Ag(111) interface by the segregation growth of germanane. A simple annealing process in $N_2$ or $H_2$/Ar at ambient pressure led to the formation of germanene, indicating that a UHV condition is not necessary. Owing to the chemical protection by the top vdW material, the grown germanene at the interface is stable in air. We found that it is necessary to use a vdW material as a cap layer for the present germanene growth method since the use of an $Al_2O_3$ cap layer resulted in the absence of germanene formation. We also proved that Raman spectroscopy in air is a powerful tool for characterizing germanene at an interface. The direct growth of h-BN-capped germanene is considered to be a promising technique for the fabrication of future germanene-based FETs.

## 4. Experimental Section/Methods

*Ag deposition:* A Ge(111) surface was chemically cleaned by dipping in diluted HF for 1 ~ 3 min, then rinsed in deionized water, and dried by blowing $N_2$. Subsequently, a Ag thin film of 70 nm thickness was deposited at RT by an e-beam evaporator (RDEC Co., with the Ltd.,

RDEB-1206K). The deposition rate was 0.4 Å $s^{-1}$. The deposited Ag film was aligned with the (111) crystal direction by solid-phase epitaxy on Ge(111).[7a]

*Transfer of CVD graphene:* CVD graphene was grown on a commercial Cu sheet (100 μm thickness, Nilaco) in a laboratory-made CVD system in the National Institute for Materials Science. The growth of polycrystalline monolayer graphene with high coverage was targeted. The typical growth process was as follows: (i) increasing the temperature from RT to 1000 ºC in Ar; (ii) $H_2$ annealing at 1000 ºC for 100 min; (iii) graphene growth at 1000 ºC by introducing methane with $H_2$ and Ar for 90 min; (iv) cooling to RT in $H_2$/Ar. Details of the growth can be found elsewhere.[31] The grown graphene was transferred onto a Ag surface by a standard wet transfer using spin-coated poly(methyl methacrylate) (PMMA) as a mechanical support, $(NH_4)_2S_2O_8$ as the etchant for Cu, and acetone to remove the PMMA layer.[20b, 32]

*Transfer of exfoliated graphene/h-BN and h-BN:* Graphene and h-BN flakes were placed on $SiO_2$/Si substrates by mechanical exfoliation using adhesive tapes. A dry transfer onto a Ag surface was conducted using laboratory-built transfer equipment and a viscoelastic polymer. Details are given elsewhere.[33]

*$Al_2O_3$ cap layer:* An $Al_2O_3$ thin film was directly deposited on a Ag surface by radio frequency sputtering (Shibaura Mechatronics Corporation, CFS-4EP-LL) under Ar and $O_2$ flows (both at 10 standard cubic centimeters: sccm) at a pressure of 0.3 Pa. The deposition rate was 0.14 Å $s^{-1}$.

*Growth of germanene at the vdW material/Ag(111) interface:* Germanane was grown at the vdW material/Ag(111) interface by annealing using a commercial rapid annealing furnace (Ulvac, RTA-6) under $N_2$ or $H_2$/Ar flow. The typical annealing sequence used was as follows: (i) increasing the temperature from RT to 250 ºC in 2 min; (ii) annealing at 250 ºC for 10 min; (iii) increasing the temperature from 250 to 550 ºC in 2 min; (iv) annealing at 550 ºC for 10 min; (v) natural cooling to RT. The role of the annealing in (ii) is to promote the crystallization of Ag(111). Since the segregation of Ge by heating at 250 ºC was negligible (**Figure S3**), the annealing in (ii) was carried out at 250 ºC.

*Characterization:* A commercial Raman microscope (Horiba, LabRam HR Evolution) with an excitation wavelength of 458 nm in a backscattering geometry was used for obtaining Raman spectra in air. Angle-resolved XPS (ULVAC-PHI, Quantera SXM) was used for surface elemental analyses. Charge correction was performed by shifting all peaks to the position of C 1s at 285.0 eV.

*In situ growth and measurement of germanene/Ag(111):* We used UHV surface analysis system consisting of two chambers. One chamber was equipped with evaporators of silver and boron, a sample heater, and AES and RHEED equipment. The other chamber contained HREELS equipment and was connected to the first chamber with a load-lock system. The sample temperature was monitored using an optical pyrometer without calibration.

*Computational calculations:* We adopted QUANTUM ESPRESSO 6.5 code,[34] the Perdew–Burke–Ernzerhof functional[35] for the exchange correlation, and the ultrasoft pseudopotentials (Ge.pbe-n-kjpaw_psl.1.0.0.UPF and Ag.pbe-n-rrkjus_psl.1.0.0.UPF) constructed by Dal Corso.[36] Basic computational parameters for each model, such as the cutoff energies of the plane wave and charge ($E_{cut}$ and $E_{cut}{}^{\rho}$), the amounts of Brillouin zone sampling for the electrons and phonons (k point and q point), and the threshold for residual forces for atoms were set after evaluating the convergence, precision, and calculation efficiency. Tables 1 and 2 show the parameters used for structural optimization and phonon calculations, respectively.

The surface germanene was modeled with a Ag(111) (7 × 7)-unit four-layer slab supercell divided by an approximately 2-nm-thick vacuum layer. Ag atoms in the first layer, which were in contact with germanene, were movable, while the other Ag atoms were fixed at bulk positions.

Table 1. Computational parameters for structural optimization

| Material | Ecut (Ry) | Ecut$^{\rho}$ (Ry) | k point | Smearing (mRy) | Residual force for each atom (mRy/bohr) | Convergence error (Ry/cell) |
|---|---|---|---|---|---|---|
| Ge bulk | 70 | 320 | 24 × 24 × 24 | 3.75 | < 0.01 | < $10^{-11}$ |
| freestanding germanene | 70 | 320 | 24 × 24 × 1 | 3.75 | < 0.01 | < $10^{-11}$ |
| germanene on Ag (111) (7 × 7) | 30 | 320 | 1 × 1 × 1 | 7.50 | < 0.01 | < $10^{-11}$ |

Table 2. Computational parameters for phonon calculations

| Material | Ecut (Ry) | Ecut$^{\rho}$ (Ry) | k point | Smearing (mRy) | q point | Convergence error (Ry/cell) |
|---|---|---|---|---|---|---|
| Ge bulk | 70 | 320 | 12 × 12 × 12 | 7.50 | 4 × 4 × 4 | < $10^{-11}$ |
| freestanding germanene | 70 | 320 | 36 × 36 × 1 | 2.50 | 8 × 8 × 1 | < $10^{-11}$ |

**Acknowledgements**
This study was supported in part by JSPS KAKENHI Grant number 17K18224 (Grant-in-Aid for Young Scientists (B)) from the Ministry of Education, Culture, Sports, Science and Technology (MEXT), Japan; NIMS Nanofabrication Platform supported by "Nanotechnology Platform Program" of the MEXT, Japan, Grant Number JPMXP09F19NMN010; Namiki Foundry in NIMS, Japan; Materials Analysis Station in NIMS; NIMS TEM station; the Public/Private R&D Investment Strategic Expansion Program (PRISM) from Cabinet Office, Japan; the Center for Functional Sensor & Actuator (CFSN) from NIMS. The DFT and DFPT calculations in this study were performed on Numerical Materials Simulator at NIMS.

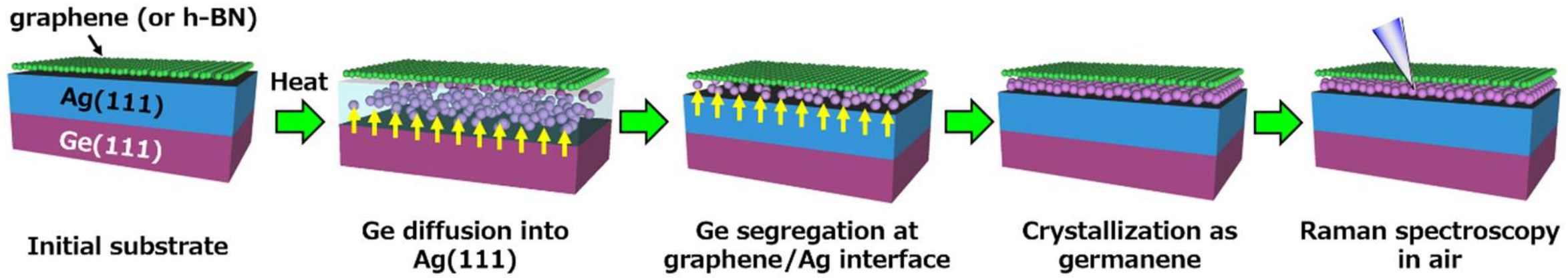

**Figure 1.** Illustration of germanene growth at vdW material/Ag (111) interface and its characterization by Raman spectroscopy in air.

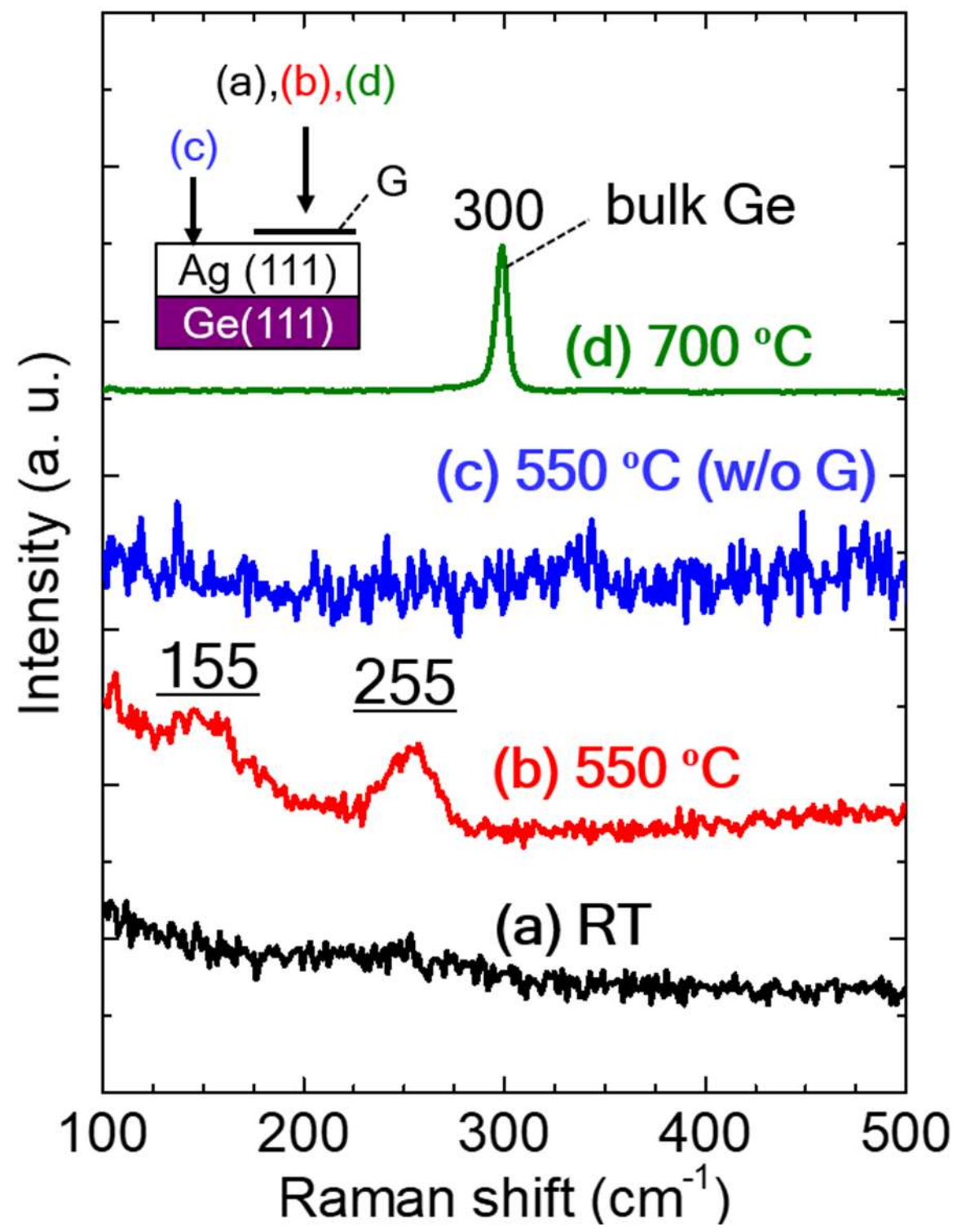

**Figure 2.** Raman spectra of graphene(G)/Ag(111)/Ge(111) after annealing at (a) RT, (b) 550, and (d) 700 ºC, and (c) Raman spectrum of Ag(111)/Ge(111) without graphene after annealing at 550 ºC. The top graphene was placed onto the Ag surface by the wet transfer of CVD graphene. The inset shows the areas where the spectra were obtained.

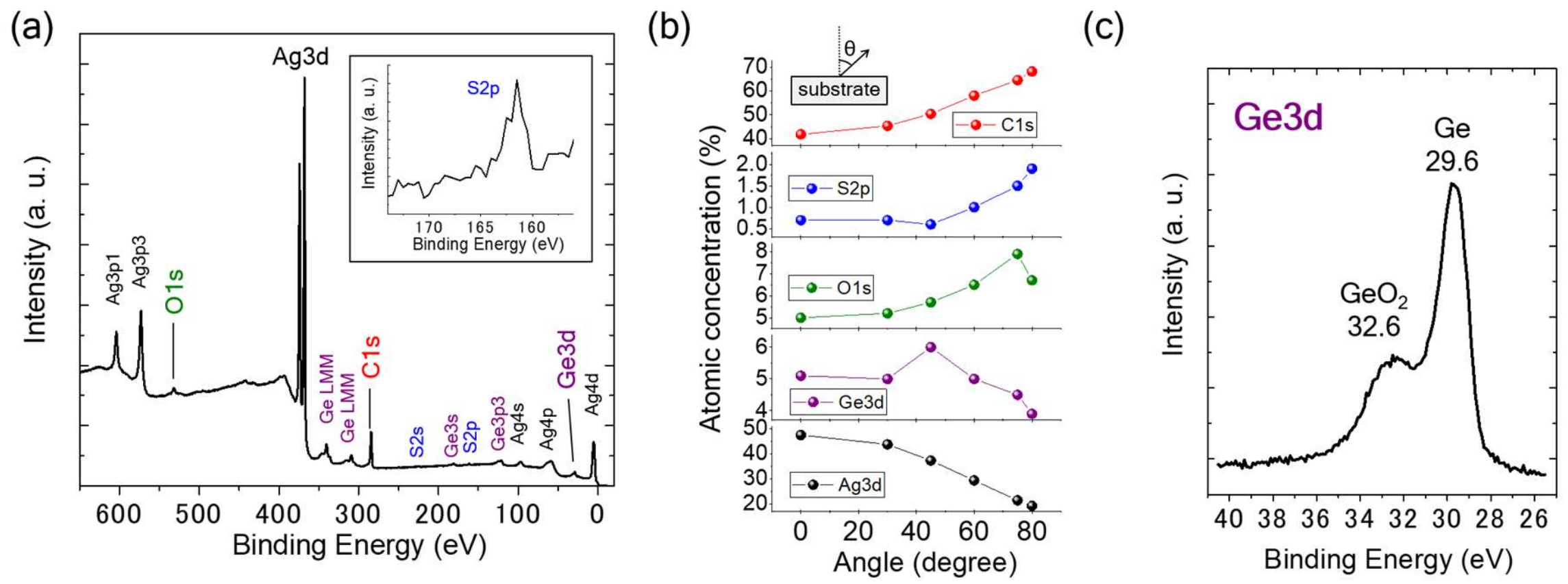

**Figure 3.** (a) XPS spectrum of graphene/Ag(111)/Ge(111) annealed at 550 ºC. (b) Atomic concentration of each element as a function of the photoelectron detection angle. The inset shows the definition of the detection angle. (c) Ge 3d peak obtained at 45º with high energy resolution. The annealed sample was kept in air for more than 42 days before introducing it into the XPS chamber.

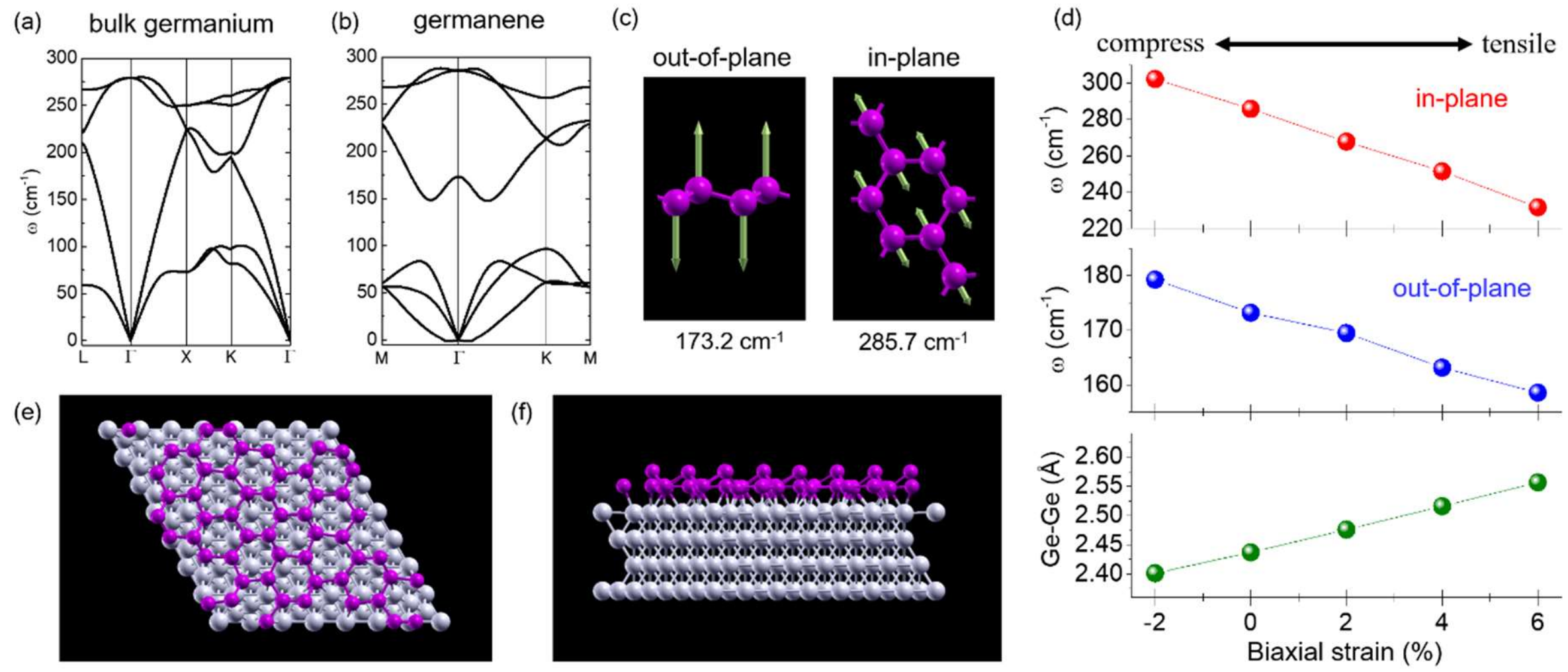

**Figure 4.** DFT calculations. Phonon dispersions of (a) bulk Ge and (b) freestanding germanene. (c) Illustration of Raman active vibration modes in germanene. (d) Phonon energies of in-plane and out-of-plane modes and the Ge-Ge bond length in freestanding germanene as a function of in-plane biaxial strain. (e) Top and (f) side views of the calculated germanene structure on four-layer Ag(111). Structural optimization was carried out with movable Ge and Ag atoms in the first surface layer and fixed Ag bulk positional atoms in the second to fourth surface layers. The average Ge-Ge bond length in germanene on Ag(111) was ~ 2.49 Å.

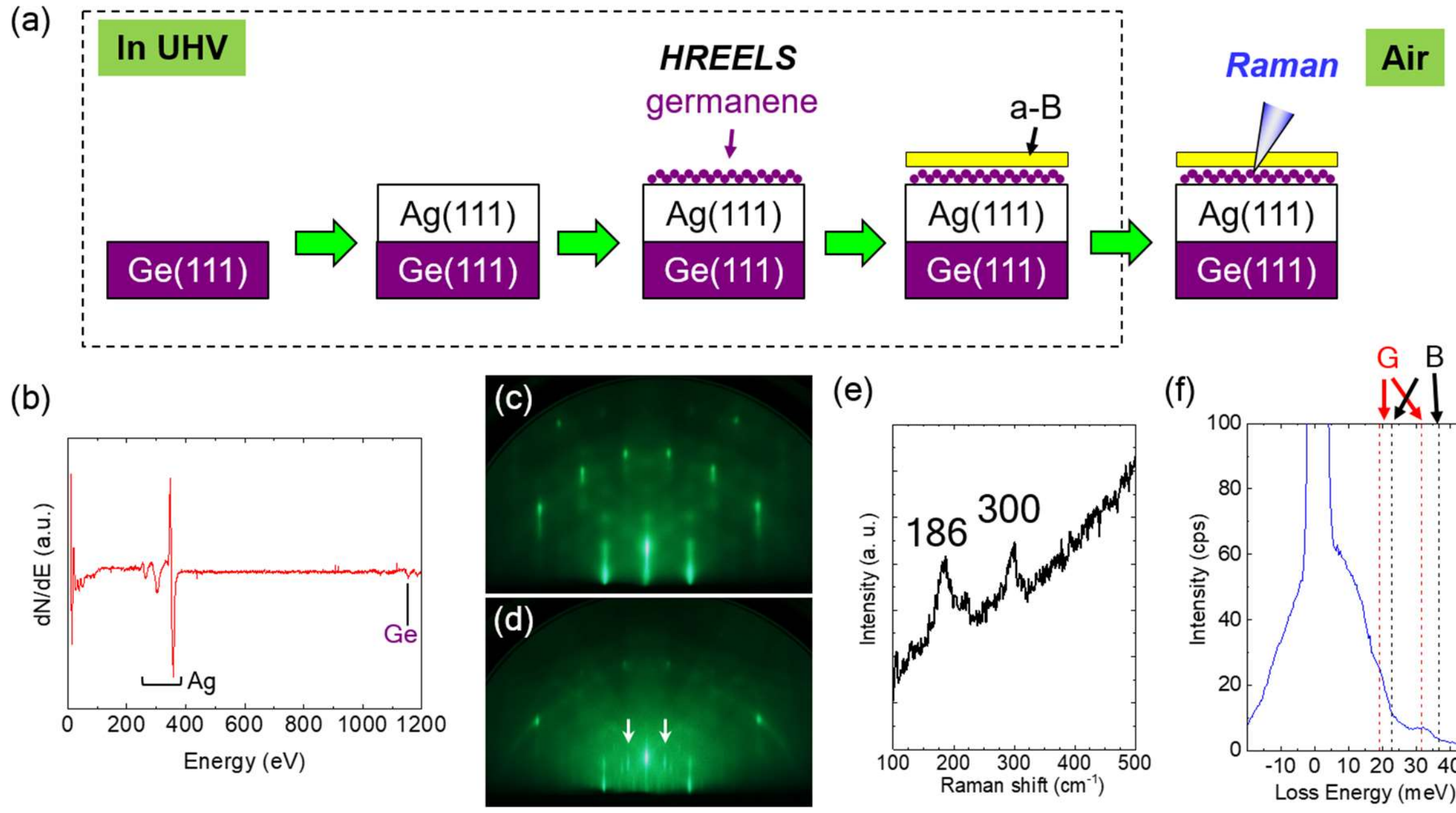

**Figure 5.** *In situ* growth of germanene. (a) Illustration of experimental procedure. (b) AES spectrum of germanene on Ag(111). RHEED patterns of (c) Ag(111) and (d) germanene/Ag(111) surface. Streaks with arrows correspond to germanene. The azimuth of the incident e-beam is $[11\bar{2}0]$. (e) Raman spectrum of boron-coated germanene on Ag(111) recorded in air. (f) Off specular HREEL spectrum of germanene on Ag(111). Black (B) and red (G) dotted lines show the phonon energies of germanene at the graphene/Ag(111) and boron/Ag(111) interfaces, respectively, which were obtained by Raman spectroscopy.

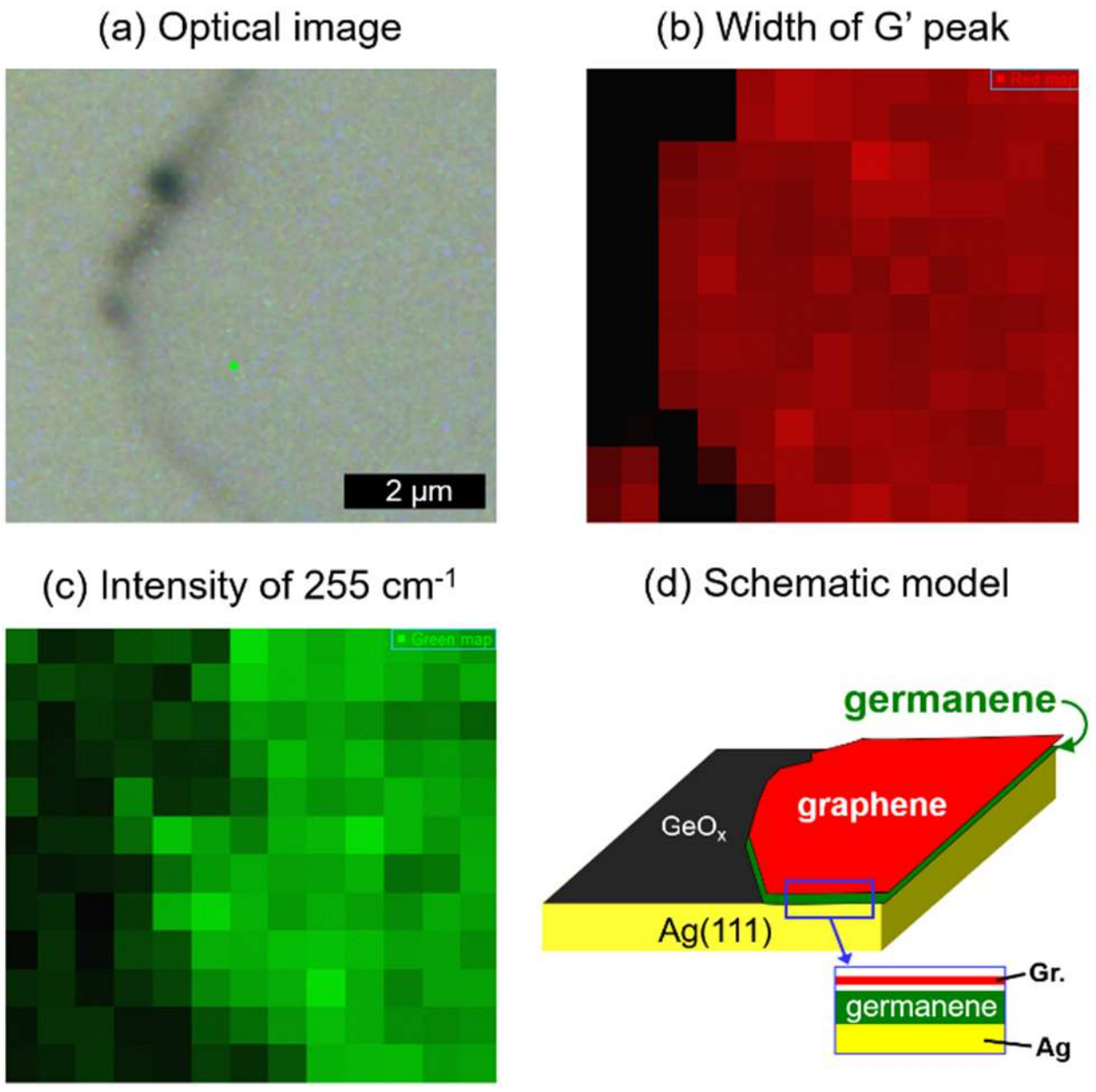

**Figure 6.** Raman mapping of germanene at graphene/Ag(111) interface. (a) Optical microscopy image of the sample. Raman maps for (b) full width at half maximum (FWHM) of the G' peak (graphene) and (c) peak intensity at 255 $cm^{-1}$ (germanene). (d) Schematic model of the Raman-mapped area of the sample.

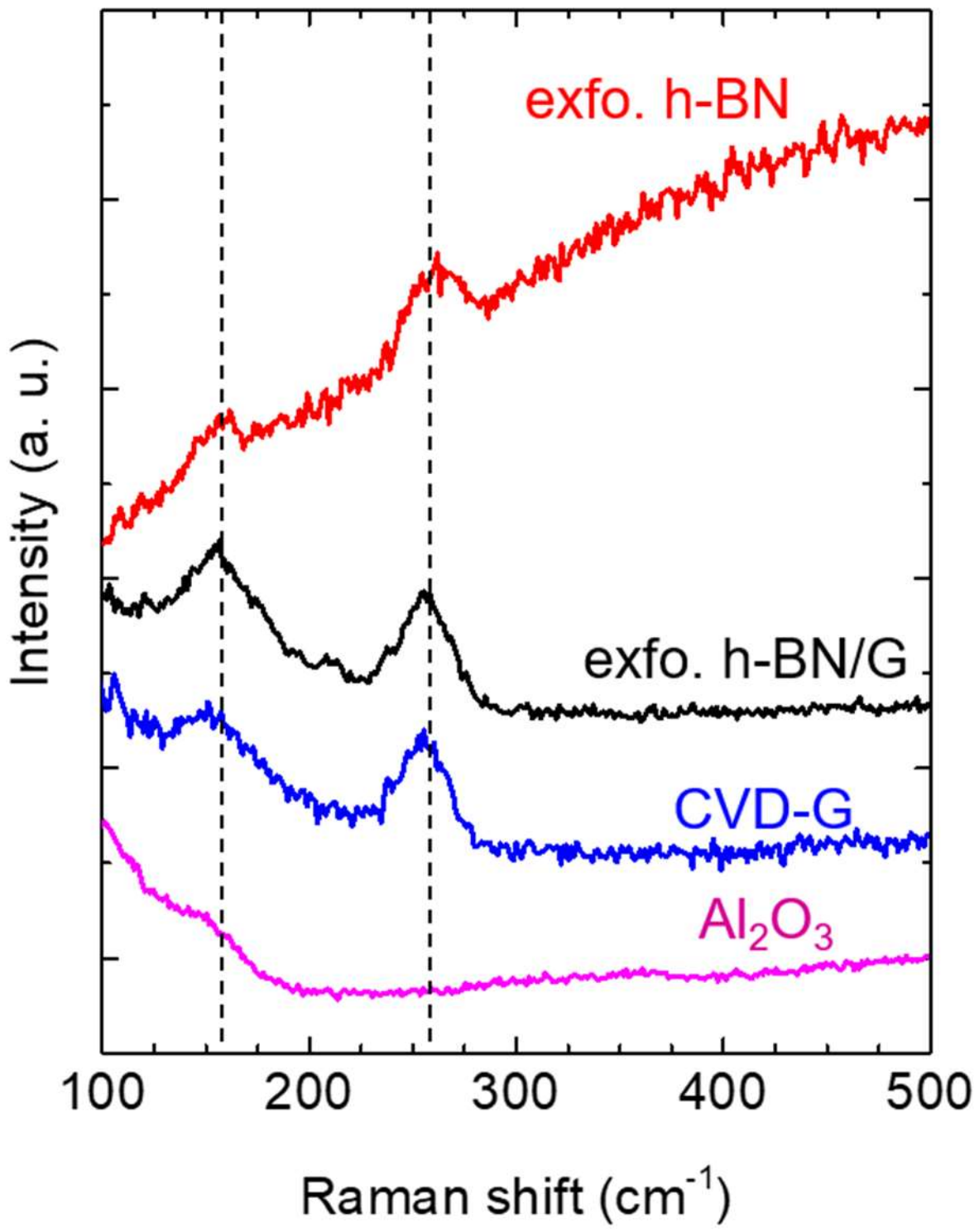

**Figure 7.** Raman spectra of germanene with the different cap layers. Multilayer exfoliated h-BN (~ 30 nm thickness), exfoliated graphene (monolayer) stacked with h-BN, CVD graphene (monolayer), and sputtered $Al_2O_3$ (~20 nm thickness) were deposited on Ag(111)/Ge(111). The annealing conditions were 550 ºC under $N_2$ ambient.

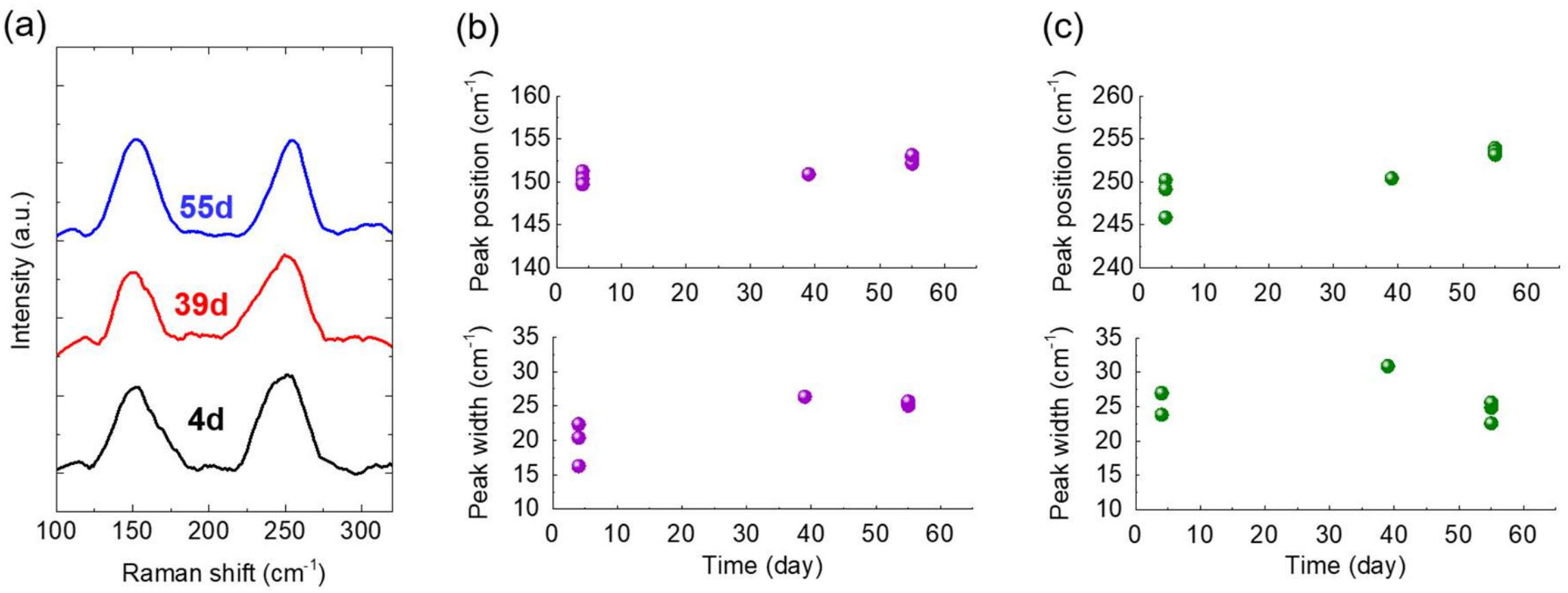

**Figure 8.** (a) Raman spectra of germanene at the h-BN/Ag(111) interface after keeping it in air for 4, 39, and 55 days. Smoothing and subtraction of baselines have been performed for the spectra, and then the peaks were fitted by a Lorentzian curve. Extracted peak positions and FWHM of (b) out-of-plane and (c) in-plane vibration modes of germanene.

**Table 1.** Computational parameters for structural optimization.

| Material | $E_{cut}$ (Ry) | $E_{cut}^{\rho}$ (Ry) | k point | Smearing (mRy) | Residual force for each atom (mRy/bohr) | Convergence error (Ry/cell) |
|---|---|---|---|---|---|---|
| Ge bulk | 70 | 320 | 24 × 24 × 24 | 3.75 | < 0.01 | < $10^{-11}$ |
| freestanding germanene | 70 | 320 | 24 × 24 × 1 | 3.75 | < 0.01 | < $10^{-11}$ |
| germanene on Ag (111) (7 × 7) | 30 | 320 | 1 × 1 × 1 | 7.50 | < 0.01 | < $10^{-11}$ |

**Table 2.** Computational parameters for phonon calculations.

| Material | $E_{cut}$ (Ry) | $E_{cut}^{\rho}$ (Ry) | k point | Smearing (mRy) | q point | Convergence error (Ry/cell) |
|---|---|---|---|---|---|---|
| Ge bulk | 70 | 320 | 12 × 12 × 12 | 7.50 | 4 × 4 × 4 | < $10^{-11}$ |
| freestanding germanene | 70 | 320 | 36 × 36 × 1 | 2.50 | 8 × 8 × 1 | < $10^{-11}$ |

**Direct growth method of germanene at interfaces between vdW materials and Ag(111)** is proposed and developed. The grown germanene is stable in air, which enables its handling in air. A vdW interface provides a nanoscale platform for growing germanene similarly to that in vacuum, which cannot be achieved with a typical oxide interface such as $Al_2O_3$.

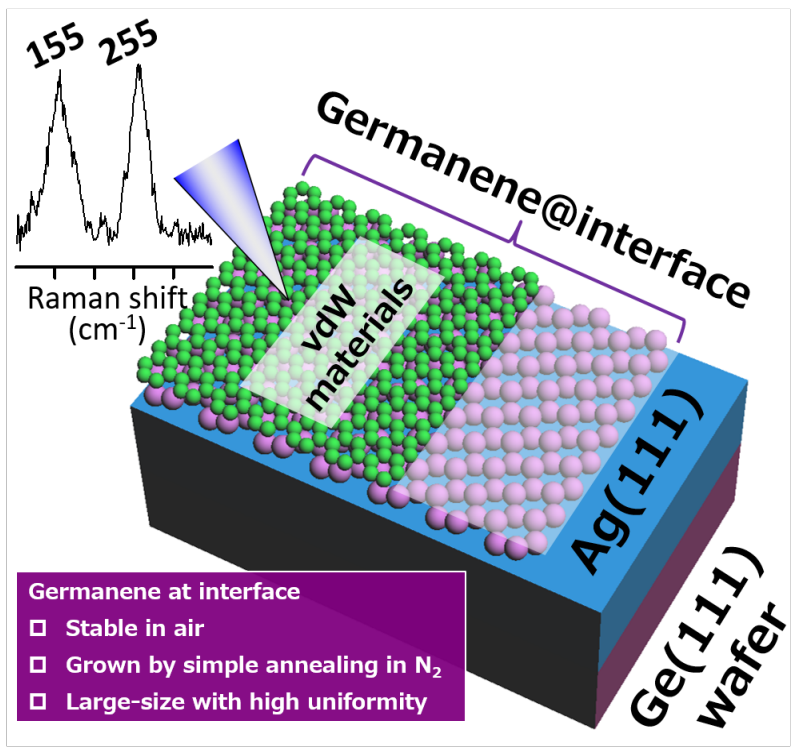

# Supporting Information

**Direct growth of germanene at interfaces between van der Waals materials and Ag(111)**

*Seiya Suzuki,* Takuya Iwasaki, K. Kanishka H. De Silva, Shigeru Suehara, Kenji Watanabe, Takashi Taniguchi, Satoshi Moriyama, Masamichi Yoshimura, Takashi Aizawa, and Tomonobu Nakayama*

## S1. Raman spectra of annealed graphene/Ag(111)/Ge(111)

**Figure S1**a and Figure S1b are, respectively, magnified and whole Raman spectra of annealed CVD graphene/Ag(111)/Ge(111) in $H_2$/Ar. Germanene peaks (155 and 255 $cm^{-1}$) were observed from 450 to 600 ºC. High temperature annealing resulted in increasing D peak which indicates creation of defects in graphene.

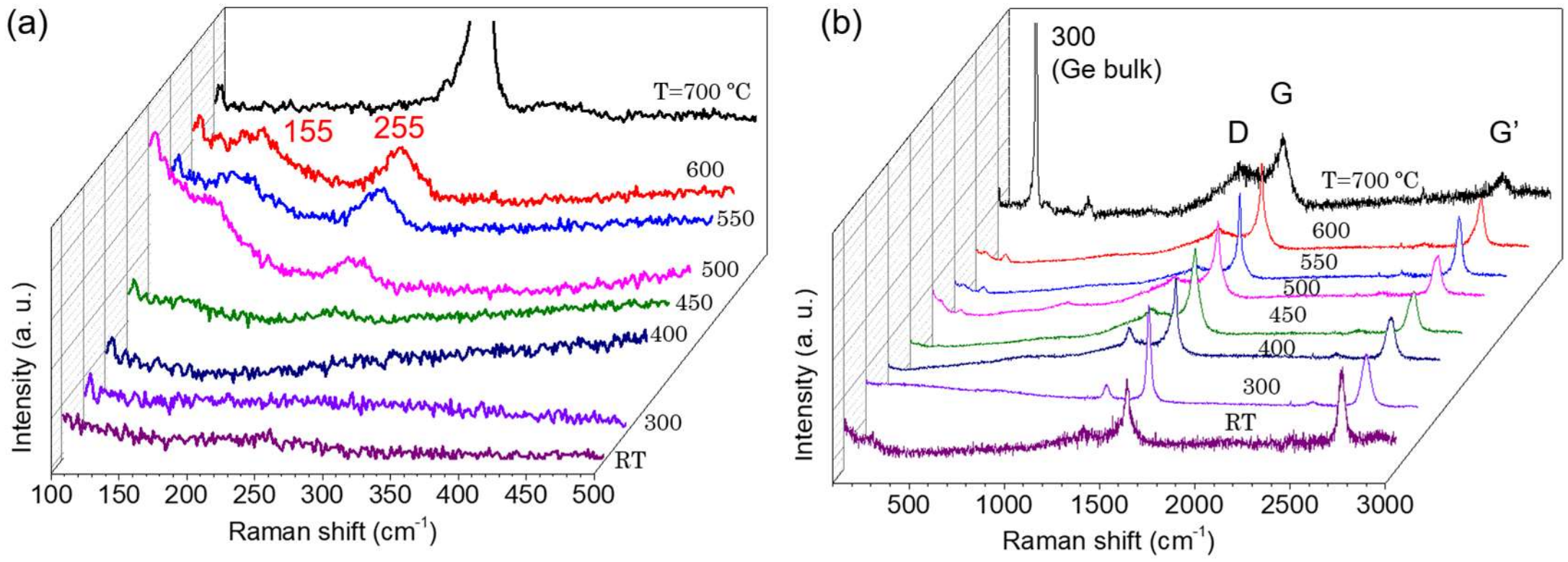

**Figure S1.** (a) Magnified and (b) whole Raman spectra of annealed CVD graphene/Ag(111)/Ge(111) in $H_2$/Ar.

**S2. Cross-sectional scanning transmission electron microcopy (STEM)**

To examine the layer number of germanene, we performed cross-sectional high-angle annular dark field (HAADF)-STEM for the annealed graphene/Ag(111)/Ge(111) sample at 550 ºC. **Figure S2**a shows a cross-sectional Z-contrast image of the sample around the graphene/Ag(111) interface. In Z-contrast images, the contrast is proportional to the square of the atomic number. The rectangel lattice, which composed of bright atoms in Figure S2a, corresponds to Ag (111).

To detect germanene, we perormed HAADF-STEM and energy dispersive X-ray spectroscopy (EDS) analysis with the limited scanning area in the green rectangle in Figure S2a. The layer number was named as 1 ~ 10 from the interface side to Ag (111) bulk side of the sample. Figure S2b-d show the Z-contrast and the EDS singnals of Ag and Ge. Figure S2e shows the overlapped EDS signals of Figure S2c and FigureS2d. Figure S2f shows the plots of the normalized average signals of the Z-constrast, the EDS Ag, and the EDS Ge. Periodic signals of Z-contrast and EDS Ag were observed for 3 ~ 10 layers, corresponding to Ag (111) crystal planes. The EDS Ge singnals were slightly higher at the $1^{st}$ layer than the other. This indicates that germanene layer is probabely monolayer not multilayer.

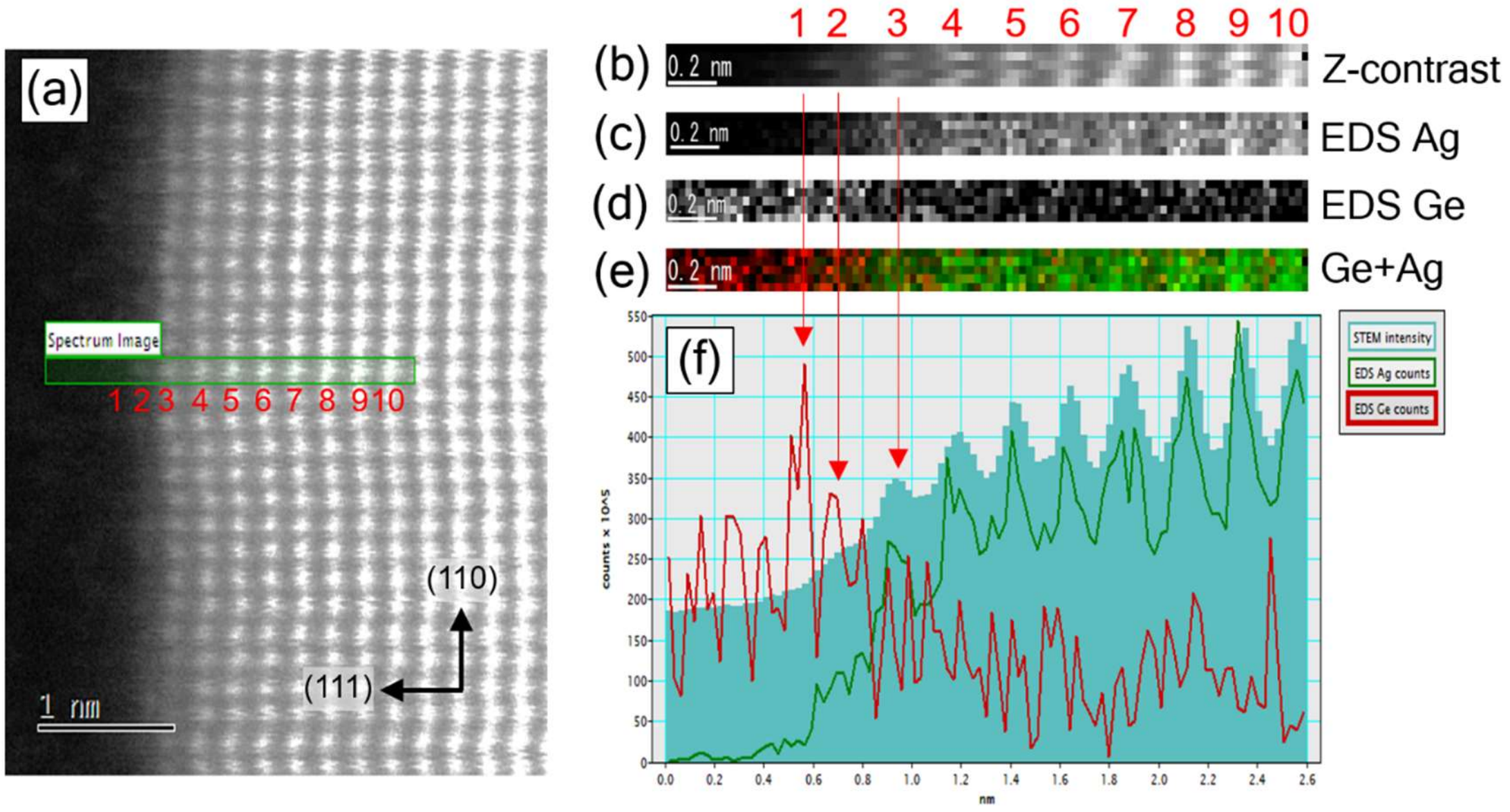

**Figure S2.** Cross-sectional STEM observation of the graphene/Ag(111)/Ge(111) annealed at 550 ºC. (a) Cross-sectional Z-contrast image around the graphene/Ag(111) interface. The arrows in (a) indicates the crystal direction of Ag. (b) Z-contrast and EDS singnals of (c) Ag and (d) Ge in the green rectangle in (a). (e) Overlapped EDS signals of (c) and (d). Red and green color connrespond to Ge and Ag, respecively. (f) Plots of the normalized average signals of the Z-constrast (filled green), the EDS Ag (green line), and the EDS Ge (red line).

### S3. Segregation of Ge from Ag(111)/Ge(111)

**Figure S3** shows the temperature dependence of the surface atomic ratio of Ge/Ag of Ag(111)/Ge(111) substrate. The Ge/Ag ratios were calculated from the peak area of Ge 3d and Ag 3d, which are obtained by XPS, with taking into account the photoionization cross-section for each element. Blue and red dots show same data but in different scale of Ge/Ag. Segregation of Ge was started from 300 °C. Ge/Ag was increased until 350 °C and remained nearly constant at 0.3 until 650 °C (see blue dots). Further heating led the drastic increase of Ge segregation and the Ge/Ag ratio reached ~ 12 (see red dots). The drastic increase of Ge/Ag is due to the disappearance of the Ag layer by its phase transition to liquid since 700 °C is sufficiently higher than the eutectic melting point of Ag-Ge system (~ 650 °C).

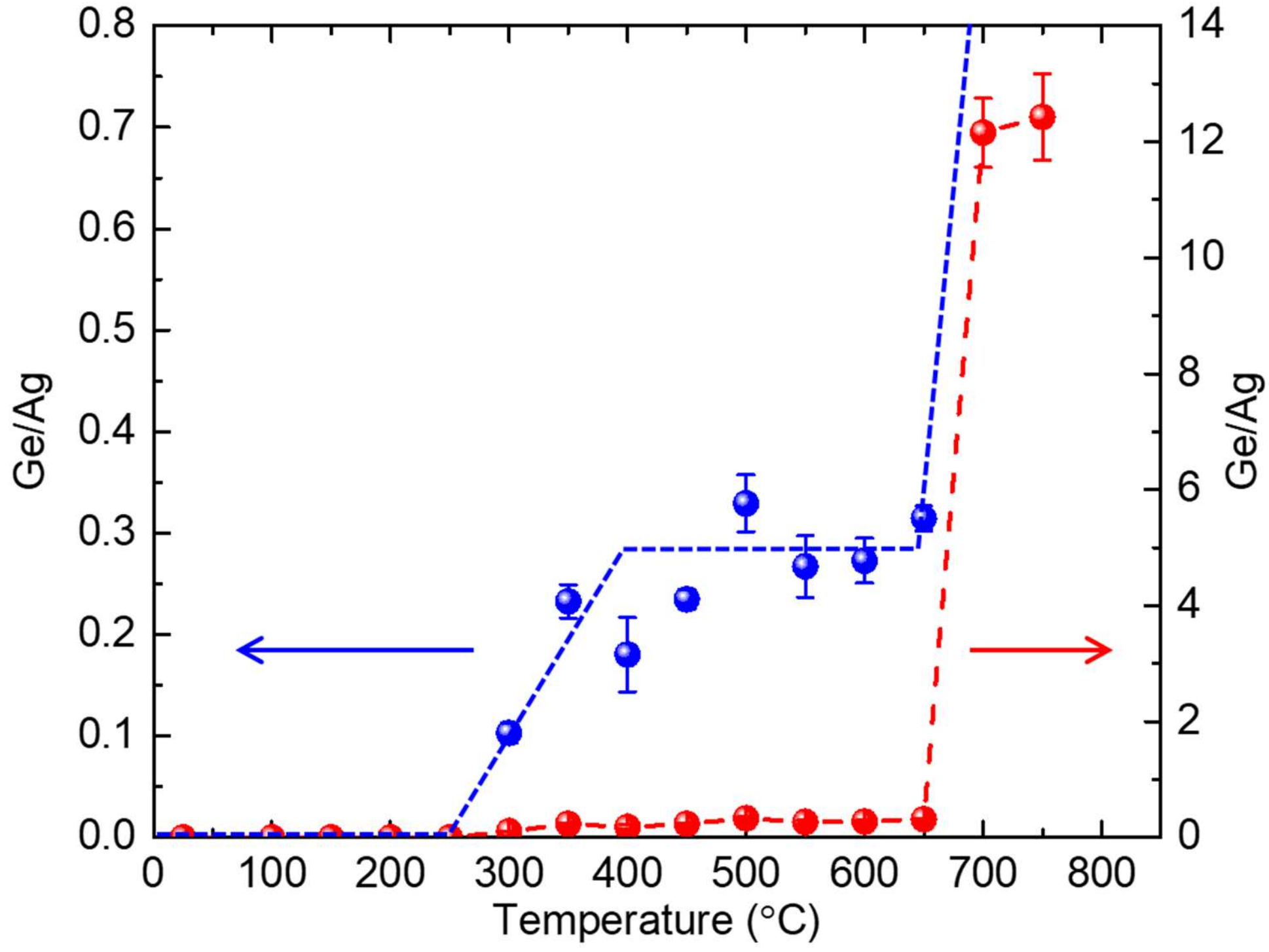

**Figure S3.** Temperature dependence of the surface atomic ratio of Ge/Ag of Ag(111)/Ge(111) substrate. Blue and red dots show same data but in different scale of Ge/Ag.